# Environmental effects on the spread of the Neolithic

K. Davison[1*], P. M. Dolukhanov[2], G. R. Sarson[1] and A. Shukurov[1]

**Abstract** The causes and implications of the regional variations in the spread of the incipient agriculture in Europe remain poorly understood. We apply population dynamics models to study the dispersal of the Neolithic in Europe from a localized area in the Near East, solving the two-dimensional reaction-diffusion equation on a spherical surface. We focus on the role of major river paths and coastlines in the advance of farming to model the rapid advances of the Linear Pottery (LBK) and the Impressed Ware traditions along the Danube-Rhine corridor and the Mediterranean coastline respectively. We argue that the random walk of individuals, which results in diffusion of the population, can be anisotropic in those areas. The standard reaction-diffusion equation is thus supplemented with advection-like terms confined to the proximity of major rivers and coastlines. The model allows for the spatial variation in both the human mobility (diffusivity) and the carrying capacity of landscapes, reflecting the local altitude and latitude. This approach can easily be generalised to include other environmental factors, such as the bioproductivity of landscapes. Our model successfully accounts for the regional variations in the spread of the Neolithic, consistent with the radiocarbon dated data, and reproduces a time delay in the spread of farming to the Eastern Europe and Scandinavia.

Keywords: NEOLITHIC, POPULATION DYNAMICS, ENVIRONMENTAL EFFECTS, EXPANSION RATE, EUROPE.

## *1. Introduction*

The transition from hunter-gathering to early forms of agriculture and stock breeding, when humanity entered the epoch of the Neolithic, was one of the leaps in human prehistory that ultimately shaped all later civilisations. The nature of the Neolithic and the mechanism for the spread of agriculture over vast areas in the Old World remain topical issues in archaeology and related disciplines, [25, 32, 30].

One viewpoint, first proposed and advocated by Gordon Childe [11], treats the spread of agriculture in Europe as a result of demic expansion, wherein farmers immigrated to Europe from the Near East, bringing with them new technologies and subsistence strategies. An alternative approach attaches more importance to cultural diffusion; i.e., the adoption of cultural traits not necessarily associated with long-range travel of individuals [42]. Despite their fundamental difference, both processes in fact represent gradual spread driven by individual random events, either human migrations or cultural exchange events. Therefore, both processes can be modelled with (almost) the same mathematical equations involving the diffusion operator (or its appropriate generalization), albeit with different parameters.

---

[1] School of Mathematics and Statistics, University of Newcastle upon Tyne, NE1 7RU, U.K.
[*] (kate.davison@ncl.ac.uk)
[2] School of Historical Studies, University of Newcastle upon Tyne, NE1 7RU, U.K.

It is appealing to apply population dynamics models to describe quantitatively the spread of the Neolithic. The simplest model of this type was suggested by Ammerman and Cavalli-Sforza [3], who chose parameter values appropriate for demic expansion. This model neglected any heterogeneity of the environment (and only suggested a mild latitudinal gradient in the rate of spread); even coastlines were neglected at that level of approximation. Nevertheless, the model was remarkably successful in explaining the constant rate of spread of incipient farming over the vast area from the Near East to Western Europe. The constant speed of front propagation is a salient feature of solutions to one of the most popular equations of population dynamics, the Fisher–Kolmogorov–Petrovsky–Piskunov (FKPP) equation (e.g. [19,20]) in one dimension. Further developments of this model make clear the need to include heterogeneity of the areas where the population spreads (e.g., [36]); that is the aim of this paper. We discuss regional variations in the spread of the Neolithic in Europe, most notably the rapid advances of the Linear Pottery (LBK) and the Impressed Ware traditions along the Danube–Rhine corridor and the Mediterranean coastline, respectively. In this paper, we focus on the role of major river paths and coastlines in the advance of farming. We argue that the standard reaction-diffusion equation for the population density must be supplemented with an advection term confined to the proximity of rivers and coastlines. We show that this significantly affects the *global* propagation rate of agriculture, even when the advection is restricted to the narrow vicinity of water pathways. Furthermore, our model includes altitude above the sea level, together with mild latitudinal variations, as additional important environmental parameters.

## *2. The Neolithic expansion*

### 2.1. Archaeological evidence

The origins of European agriculture are normally sought in the Near East. The earliest indications of agriculture, in the form of cultivation of cereals and pulses, and rearing of animals, come from the Zagros foothills. Their age, 11,700–8,400 BP [8], corresponds to the cool, dry climatic period (the Younger Dryas) followed by a rapid increase in rainfall at the beginning of the Holocene (ca. 10,000 BP).

During the early stages of agricultural development (the Preceramic Neolithic A and B, 9800–7500 cal BC), the rapid increase in the number of sites is noticeable in both the foothills and the surrounding plains, accompanied by the appearance of large settlements with complicated masonry structures and fortifications (e.g. Jericho).

At a later stage, the core area of early agricultural settlements shifts to the north, to the eastern highlands and inner depressions of Asia Minor. The most outstanding case of early agricultural development in this area is Çatal Hüyük, a Neolithic town on the Konya Plain (ca. 6500–5700 cal BC) [7].

The earliest sites with developed agricultural economies in Europe, dated 6400–6000 cal BC, are found in the intermontane depressions of Greece (Thessaly, Beotia and Pelloponnese). Genetic features of the cultigens and the general character of the material culture leave no doubt to their Near-Eastern origins. Significantly, the early Neolithic sites in the Marmara Sea basin (the Firiktepe culture) are of a more recent age (6100–5600 cal BC), being culturally distinct from the Early Neolithic in Greece. This implies that the Neolithic communities could penetrate the Balkan Peninsula from Western Asia by means of navigation.

The Neolithic spread further, plausibly via the Strouma axis in the northeast, and the Vardar–Morava axis in the north. The ensuing development saw a rapid growth of Neolithic settlements in the depressions of northern Thrace, the Lower and Middle Danube catchment basin (the cultural complex Karanovo-Starcevo-Koros-Cris, KSKC, 5900–5500 cal BC).

The next stage in the Neolithic development saw the emergence of early LBK sites in the form of KSKC-type settlements on the Tisza Plain in 5600–5500 cal BC. [9].

The LBK sites later spread over the vast areas of the loess plains of Central Europe, mostly along the Danube, Rhine–Mainz and Vistula axes. This spread occurred within the range of 5600–4800 cal BC, with the most probable age of 5154±62 [17].

The spread of early agricultural communities further east, into the forest–steppe areas of the East European Plain, is evinced by the Cucuteni–Tripolye sites, of age 5700–4400 cal BC [10].

All the aforementioned cultural entities bear cultural affiliations with the early agricultural communities of Western Asia, implying that the spread stemmed from that area.

Recent research also identifies a different pattern of the Neolithisation in Europe, less obviously related to the Near East. Numerous pottery-bearing sites have been found along the Mediterranean coastal areas of France and Spain, as well as in the Atlantic coastal regions of France and Portugal. These sites, referred to as Epi-Cardial and Roucadour, show an early age of 7350–6500 and 6400–5500 cal BC, respectively. They are often viewed as related to the La Hoguette sites in north-eastern France, Belgium and north-western Germany, which apparently predate the LBK tradition [40].

There also exists convincing evidence of early pottery making on the East European Plain, related to such cultural entities as Yelshanian (6910 ± 58 cal BC), Buh-Dniestrian (6121 ± 100 cal BC), Rakushechnyi Yar (5846 ± 128 cal BC), and early pottery-bearing cultures in the forested areas (5317 ± 30 cal BC) [17]. This evidence reveals a Neolithic stratum which apparently predated the Near-Eastern 'wave of advance', and later interacted with it.

## 2.2 Demic expansion versus cultural transmission

Demographic pressure is generally considered to be the prime mover of the Neolithic expansion. The Ice Age hunter-gathering groups existed in an equilibrium eco-social system; they were able to control their population in response to variations in food supply. Thus, births are normally spaced at 3–5 year intervals among *!Kung* nomadic hunter-gatherers in South Africa, where the maximum potential fertility per woman is reduced to 3–5 children and often further diminished by infanticide and high mortality [27].

Judging from the number of sites, the population in the Near East started increasing ca. 15,000 BP with the appearance of Natufian sites, marked by an increased sedentariness and a broadened range of subsistence strategies. Apparently, the natality rate dramatically increased with the emergence of agricultural sedentary settlements. This was supposedly related to the changed social status of women and to better childcare, combined with a more nutritious food supply. The rapid population growth was partly hampered by the adverse effect of higher population density, which caused higher risk of infectious diseases. Yet several Neolithic agglomerations already reached impressive sizes at the aceramic stage; e.g., Jericho A, with a population of 2000–3000, and Çatal Hüyük, with ca. 8000 inhabitants [26].

Since Childe [11], mass migration from western Asia was deemed as the most viable mechanism of Neolithic expansion into Europe. More recent studies [42, 31] attach greater significance to the indigenous adoption of agriculture, described as cultural diffusion, driven by contacts between invading farmers and local foragers. It is clear, however, that some human migration occurred at each stage of the Neolithisation. First, molecular genetic evidence convincingly proves the Near-Eastern origins of the major domestic animals (sheep, goat, cattle and pig) and plants (barley, wheat and pulses) [38, 25]. Second, the human DNA demonstrates that at least 10–15% of the existing genetic lineages were introduced into Europe in the course of the Neolithisation from the Near East [33]. Third, early Neolithic archaeological assemblages in south-eastern and central Europe have little or no common elements with the preceding Mesolithic cultures, implying the influx of new populations.

On the other hand, there is sufficient evidence that groups of hunter-gatherers were variably involved in the process of Neolithisation. This is suggested by the occurrence of Mesolithic-type lithic tools in several early agricultural assemblages (in some areas of the LBK and in Tripolye). Strontium-isotope analysis of skeletal remains at several LBK sites in the Rhine Valley strongly suggests intermarriages between farmers and hunter-gatherers occurred [24]. Significantly, both direct migration and cultural diffusion resulted in a significant population growth.

Discussing the Neolithic expansion, one should consider several important environmental constraints. All early farming sites were located in areas with fertile and easily arable soils, and in close proximity to water reservoirs (lakes or rivers). Mixed broad-leaved forests with natural clearances were evidently favoured by early farmers. The natural habitats of early agricultural settlements enjoyed a considerable amount of rainfall and sufficiently high temperature during the vegetation period, which guaranteed a satisfactory yield. Significantly, the periods of major agricultural advances coincided with the periods of increased temperature and rainfall (the Holocene climatic optima).

The agricultural expansion is now often viewed as a leap-frog migration, with comparatively small groups establishing semi-permanent settlements in agricultural oases along major rivers or sea coasts, and partially budding off further afield, when the population reached a critical mass.

It is thus clear that farmers' migration into Europe did not occur in a uniform way; indeed spatial variations in the propagation speed of the land farmers have been noted [2, 23, 12]. This is not surprising when the heterogeneity of the spatial domain, Europe, is considered. The available radiocarbon dates on the spread of farming and pottery making in Northern Eurasia are summarised in Fig. 1.

## 2.3. Demographic models

The propagation of land farming throughout Europe has attracted sustained interest (see, e.g., [1] for a recent review). Edmonson [18] conducted a pioneering study into Neolithic diffusion rates. His empirically relevant hypothesis was that the apparent propagation rate of simple, rational Neolithic traits was approximately constant; he estimated it to be 1.9 km/year. This estimate refers to a far larger area than Europe. Edmonson assumed that he was measuring cultural diffusion.

Clark ([12], especially Fig. 2 there) presented a map of radiocarbon dated sites to visualise the spread of farming in Europe. Only a limited number of dates were used here, with only the date of first settlement in the area being shown. Clark supported the idea that the Neolithic penetrated Europe from the south-east, progressed along the Danube and Rhine valleys, and then throughout the rest of Europe. The Neolithic culture in Scandinavia developed much later.

Ammerman and Cavalli-Sforza [2], focused on measuring the rate of spread of early farming in Europe. In contrast to Edmonson, they based their measurement on a single trait and restricted their study to a far more specific geographical area. These authors derived the rate of spread to be $U \approx 1$ km/year on average in Europe; this estimate has remained widely accepted since then. They also noted very significant regional variations in the rate of spread. For example, unfavourable ecological and geographical factors caused a retardation of spread to the Alps; similarly retarded spread occurs at latitudes above 54º N. The Danube and Rhine valleys, the propagation path of the LBK, had an increased propagation speed, as did the Mediterranean coast [43]. The speeds of propagation of the wave front, $U$, in these areas are as follows:

$$
\begin{aligned}
&U \approx 1 \text{ km/yr} && \text{on average in Europe,} \\
&U \approx 4\text{–}6 \text{ km/yr} && \text{for the Danube–Rhine valleys,} \\
&U \approx 10 \text{ km/yr} && \text{in Mediterranean coastal regions.}
\end{aligned} \quad (1)
$$

Interpretations of these observations are usually based on the reaction-diffusion equation of population dynamics (known as the Fisher–Kolmogorov–Petrovskii–Piskuniov equation; FKPP hereafter) [19, 20]. The constant propagation speed of the population front is a salient feature of solutions to this equation in one dimension (e.g., [28]). However, applications of this approach to the spread of the Neolithic in Europe have hardly advanced beyond simple one-dimensional models in a homogeneous environment.

The results of Ammerman and Cavalli-Sforza have recently been confirmed by Gkiasta et al. [23] who used a much more comprehensive radiocarbon database. These authors suggested that the regional variations in the spread may be due to variations in the importance of demic versus cultural diffusion, with the former leading to a more abrupt transition.

While much work has been carried out into the measurement of the Neolithic dispersal, work on modelling this phenomenon is sparse. Fort and Méndez [21, 22] discuss the front propagation rate resulting from various generalizations of the FKPP equation, but their results are restricted to one dimension and to homogeneous systems. A model has yet to be developed which takes into account the influence of heterogeneous environments on the spread of farming and models it more realistically in two dimensions. Steele et al. [36] modelled the dispersal of hunter-gatherers into North America using a two-dimensional numerical model where spatial variation in the carrying capacity was allowed for (as suggested by paleovegetation reconstructions). These authors note that the diffusivity (mobility) of people must also be a function of position and time, and suggest that the spread might have followed major river valleys (see also [5]), but do not include these effects into their model.

The aim of this work is to formulate and develop a model for the spread of incipient farming in Europe, taking account of such influences; we can then investigate which environmental factors have the most significant effect. The particular environmental factors we consider are the altitude, latitude, opportunity of sea travel, major rivers and coastlines.

## 3. The Model

## 3.1 Basic Equations

We rely on the basic assumptions of population dynamics models [28, 15], which allow us to adopt the following equation for the population density *N*:

$$\frac{\partial N}{\partial t} + (\vec{V} \cdot \nabla) N = \gamma N \left[1 - \frac{N}{K}\right] + \nabla \cdot (\nu \nabla N). \quad (2)$$

This equation describes the evolution of the population density at any position, *N(θ,ϕ,t)*. Here the equation is applied at the spherical surface of the Earth, whose radius *r* is approximately 6378.1 km; *θ* is the co-latitude (*θ* =0 on the North pole) and *ϕ* the longitude. We note in this connection that deviations from planar geometry become quite pronounced on the global length scales involved, and front propagation rates inferred from planar models can be significantly in error.

The significance of the various terms in Eq. (2) is as follows. The first term on the left-hand side is the net rate of change of the population density with time at a given position. The first term on the right-hand side is the basic logistic growth term, where γ is the intrinsic birth rate (measured, e.g., in 1/year) and *K* is the carrying capacity (e.g., in persons/km²). Both γ and *K* are allowed to vary in space (and time), to model the variation in the habitats' ability to support the population. The last term on the right-hand side describes the diffusion resulting from random migration events, quantified by the diffusivity ν (e.g., in km²/year). If each member of the population is taken to move a distance λ in a random direction in two dimensions, over every time span τ (i.e. is involved in an isotropic random walk), then (see Appendix A1)

$$\nu = \frac{\lambda^2}{4\tau}. \quad (3)$$

The curve where the population density first reaches a given constant value is called the propagation front. A well-known feature of Eq. (2) in one dimension is that front propagates at a constant speed (in a homogeneous system) given by

$$U = 2\sqrt{\gamma \nu}. \quad (4)$$

In two dimensions, the propagation speed is not constant, but tends to that given by Eq. (4) as the front propagates far from its origin (so that its curvature can be neglected) [28]. Of course, the propagation speed ceases to be constant if γ and/or ν vary in space or time.

The second term on the left-hand side of Eq. (2) accounts for the enhanced motivation of the population to move in some particular directions (e.g. along river valleys); this appears in the advection velocity **V** (e.g., in km/year). Such an advection arises naturally if the random walk that underlies diffusion is anisotropic; e.g., if the length of step depends on the direction in which it is taken. If the smallest value of the random walk step is denoted λ, we can quantify the anisotropy with the parameter μ, such that the largest value of the step is given by λ(1+μ).

If the advection term were not present, all variations in the rate of spread arising from Eq. (1) must be attributed to variations in either the birth rate γ or the diffusivity ν, cf. Eq. (4). The growth rate cannot vary significantly; it is intuitively bounded from above and has been observed to be reasonably constant. Then ν must vary by a factor of 16–100 to effect variations in $U$ of order 4–10. This seems an implausible magnitude of variation, and suggests the necessity of the advection term; i.e., of anisotropic diffusion (or else of further new effects).

Simple calculations show reasonable magnitudes of anisotropy to be capable of explaining the observed accelerated spread. Substituting equation (3) into equation (4) and rearranging gives

$$\lambda = U\sqrt{\frac{\tau}{\gamma}}. \tag{5}$$

The anisotropic random walk leads to an advection speed of [15] (see Appendix A1 for derivation)

$$V = \frac{\lambda\mu}{4\tau}. \tag{6}$$

Taking τ = 15 years as an appropriate timescale, γ = 0.02 year$^{-1}$ (see Sect. 3.3) and $V$ = 5 km/year (as for the LBK rate), we require λ = 27 km and μ = 8–13. This implies that an individual may migrate in excess of 250 km over 15 yr. Although this value of the maximum migration step is rather large, it remains more plausible than the vast variation in ν required in the absence of anisotropy.

The advection velocity increases the local propagation speed, and with it the global speed of the wave of advance. The agricultural advantages of areas close to river courses [35], and the ease of travel along rivers, make travel along rivers preferred, and thereby explain the source of anisotropy in the random walk. A similar effect is clearly also plausible along coastlines.

The prescription of the components of the advective velocity deserves brief discussion here. We first identify what we call major rivers; here we consider only the Rhine and Danube in this category. We then form a quantity, $R$, which is equal to unity on the rivers and vanishes away from them. We smooth $R$ to make it differentiable, and define the unit vector components $\hat{V}_\phi$ and $\hat{V}_\theta$ (which specify the direction of **V**) by first taking $R_\theta = \frac{1}{r}\frac{\partial R}{\partial \theta}$ and $R_\phi = \frac{1}{r\sin\theta}\frac{\partial R}{\partial \phi}$, and then normalising such that

$$\hat{V}_\phi = \frac{-R_\theta}{\sqrt{R_\phi^2 + R_\theta^2}}, \qquad \hat{V}_\theta = \frac{R_\phi}{\sqrt{R_\phi^2 + R_\theta^2}}.$$

In the coastal areas, we similarly define $R$ to be unity at the coast and zero elsewhere, with similar subsequent manipulation.

The directions of advection velocity thus obtained are illustrated in Fig. 2; the magnitude of **V** will be discussed later.

## *3.2 Numerical Methods*

In order to solve Eq. (2) numerically we adopt a standard explicit Euler time stepping scheme with adaptive time step, and finite differences in the two spatial dimensions [29]. In order to maintain the stability of the explicit scheme, we choose the time step $\Delta t$ such that

$$\Delta t \leq \min\left\{ \frac{A_1}{2\nu} \frac{\Delta\phi^2 \Delta\theta^2}{\Delta\phi^2 + \Delta\theta^2}, \frac{A_2}{2V_\phi} \Delta\phi, \frac{A_3}{2V_\theta} \Delta\theta \right\}, \qquad 0 < A_i \leq 1.$$

This ensures that the population cannot move more than one grid cell in any time step.

We define a two-dimensional mesh over Europe at a resolution of 1/12°; i.e., approximately 9.27 km in the latitudinal direction, and 8.4–2.4 km in the longitudinal direction (the larger value applying at the southern edge of our domain, the smaller value at the northern edge).

The Ural Mountains (about 60ºE) form a natural eastern boundary for our pan-European simulation. Our western boundary is less important, but since we must include the most westerly reaches of Ireland and Spain, we choose 15ºW. The north and south boundaries are taken to be 75ºN and 25ºN respectively. The northern boundary, as the western one, is largely sea, while the southern boundary is taken to comfortably include Jericho (35.5ºE, 32ºN), our chosen origin of the population spread. We adopt zero flux conditions at each of the four boundaries (cf. [28]). This boundary condition does not affect any features of our solution in the north and west (where the boundary is sea), but may be important in the south and east; zero flux boundaries there are less prescriptive than the alternatives. Our simulation also requires some initial conditions; for these, we use Jericho as our point of origin [39]. We prescribe our initial population as a narrow truncated Gaussian (in two dimensions), which allows the initial population to cover an area of finite extent.

## *3.3 Model parameters*

Plausible values for the model parameters must be prescribed. For the intrinsic growth rate γ, Steele et al. [36], amongst many other authors, suggest the range γ = 0.003–0.03/year. Here we take

γ=0.02/year,

which is consistent with population doubling in 30 years. Dolukhanov [16] estimates the carrying capacity for hunter-gatherers in a region of temperate forest to be 7 persons per 100 km². Ammerman and Cavalli-Sforza [4] suggest that the carrying capacity for farmers is a factor of 50 larger, which yields

$K$ = 3.5 persons/km².

Although Europe was not all temperate forest, we here adopt this constant value; we note that, in our model, the front propagation speed is independent of $K$ (cf. Eq. 3). From Eq. (3), and taking $U$ = 1 km/year, we fix

ν = 12.5 km²/year

as the background diffusivity.

The magnitude of the advection velocity is based on the variations in $U$ given in Eq. (1). It is taken as 5 km/yr in the Danube–Rhine valleys, in the direction of advance of the population front. The advection velocity is tangent to the rivers, as described earlier, and restricted to a strip of width 20 km around the river. More precisely, the magnitude of the advection is modulated by a Gaussian envelope of half-width 10 km. The sense of the vector **V** is taken to be in the direction of locally decreasing population density. This is done by defining a vector $\hat{\mathbf{S}} = -\nabla N / |\nabla N|$, and then taking

$$\mathbf{V} = V_0 \, \mathrm{sign}(\hat{\mathbf{S}} \cdot \hat{\mathbf{V}}) \hat{\mathbf{V}},$$

with $V_0$ = 5 km/year.

Advection is similarly prescribed in the coastal areas, to model the increased propagation rate there, due to 'diffusion through sea routes' [43]. Such an accelerated advance in the coastal areas is also suggested by the rapid spread of the Impressed Ware ceramics. The contribution to the advection velocity due to the coasts is calculated similarly to that of the rivers; i.e., the velocity vectors are tangent to the coasts, and in the direction of the population front propagation. The nominal multiplier is different, however; here $V_0$ = 10 km/year.

The resulting advection velocity is illustrated in Fig. 2 for the Danube estuary, where both river and coast are shown at large scale.

The initial farming was not possible at altitudes greater than 1000 m above sea level. To reflect this, both the carrying capacity and the diffusivity are taken to decrease to zero above this height; this decrease is implemented smoothly, over a range of approximately 100 m about the cut-off height. The geographical altitude data was obtained from Ref. [37] and smoothed to the resolution of the simulations. Then both $\nu$ and $K$ are made proportional to

$$\frac{1}{2} - \frac{1}{2} \tanh\left(\frac{a - 1000\,\mathrm{m}}{100\,\mathrm{m}}\right),$$

where $a$ is the altitude in metres. This is illustrated with the shades of grey in Fig. 2, where the altitude dependence allows us to clearly see the signature of the Carpathians in the contours of diffusivity.

The impediment to travel and agriculture in the harsh climate in the north [2] is modelled by modulating $\nu$, $K$ and $V_0$ with a linear function of latitude. The resulting reduction in these parameters is by a factor of approximately 2 across the distance from Greece to Denmark.

To allow for limited sea travel, our model allows non-vanishing diffusivity in the seas; the diffusivity decreases exponentially with distance from land, over a length scale of 10 km. Both the intrinsic birth rate and the carrying capacity vanish in the sea. As a result, a weak, diffusive tail of the population can bridge relatively narrow straits (such as the English Channel or the Danish straits). This allows the British Isles to be populated (albeit with some delay), and Scandinavia to be populated via Northern Germany. This effect is illustrated in Fig. 2 (where the diffusive tail covers most of the visible area of the Aegean, for example).

## *4. Results*

Our basic model admits an arbitrarily complex set of environmental factors to be specified as functions of both position and time. In this paper, we focus on those described in Sect. 3, with the principal aim of clarifying the impact of the anisotropic random walk (advection), occurring in river valleys and coastal regions, on the global pattern of the Neolithic spread in Europe. We consider two models. In *Model 1*, we include only the logistic growth of the population and its isotropic diffusion, i.e., $\mathbf{V} = 0$ in Eq. (2), with the diffusivity ν dependent on altitude and latitude as described above. In *Model 2*, we add advection at velocity $\mathbf{V}$ along the Rhine–Danube valleys and the sea coasts. Figure 3 illustrates results obtained with Model 1. Results of our full simulation, with $V_0$ specified as described in Sect. 3, are shown in Fig. 4.

Fig. 3 shows the population front (here given by contours where the population density is 0.01 of the maximum value) at different times. Farming moves first from Jericho through Syria and Iraq and towards Turkey, while a tail of the population also travels by sea to inhabit Cyprus, which is consistent with archaeological evidence [14]. The Caucasus Mountains cause a slowing of the spread to the east of the Black Sea from the south, and instead the population propagates faster through a bottleneck at the Sea of Marmara. There is also a narrow passage on the south coast of the Black Sea, which, along with sea travel, allows a slow spread through this region. The population then spreads through European Turkey and Bulgaria, and encounters the Balkans and then the Carpathian Mountains. This mountainous region remains unpopulated and the farmers migrate around it. This behaviour of course arises from the diffusivity and carrying capacity vanishing at altitudes above 1000 m above sea level; at about 1000m a subsistence boundary [13] is established, beyond which the population is unable to spread or to be supported by the habitat.

Due to the latitudinal variation of the diffusivity, the propagation is slower where the farmers reach higher latitudes. This is most clearly illustrated at ca. 60º N in European Russia, where the average propagation speed is about 0.8 km/year, compared to 1 km/year in, say, Syria or Iraq (at 30º N). After approximately 4,000 years, the population reaches the Alps and again a subsistence boundary is formed. As farming reaches France and Belgium, the population tail extends across both the Kattegat and the Strait of Dover, thus reaching Norway and the British Isles after 5,000 and 5,500 years, respectively. Farming first enters Scandinavia by this route, rather than through Russia and Finland; this agrees with the archaeological evidence.

At a similar time, the population has reached Morocco, and crossed by sea to Gibraltar and into Spain. Iberia is populated by two distinct branches; one crossing from North Africa as described above, the other entering from Western Europe around the Pyrenees. At the final time in the simulation, the population has migrated throughout Europe and has equilibrated at the local carrying capacity.

Fig. 4 illustrates Model 2, where advection along river paths and coastlines has been included, exhibits remarkable differences from the previous model. One of the differences can be easily noticed by comparing the 500-year isochrons in Figs. 3 and 4. In Fig. 3, the population has moved, at this time, approximately 500 km into Africa; this distance is significantly increased in Fig. 4. The population again enters Europe through the bottleneck around the Sea of Marmara, but soon afterwards the enhanced mobility near the Danube takes effect, and the population moves rapidly along the river valley to form a tongue towards the north-west. This is perhaps most convincingly illustrated by the 2000-year isochron, which shows a distinct signature of the rivers. Lateral diffusion causes the population to spread out to either side of the river, and the local advection thus has a global effect on the populations advance.

Advection along the coastlines also causes the population to progress rapidly along the Black Sea shore, and thus to enter the Ukraine from the west. The advance of the population along the Rhine and Danube valleys causes the population to reach France, Belgium and Denmark sooner than in Fig. 3. The tail of the population travelling by sea across both the Strait of Dover and the Kattegat can therefore reach the British Isles and Norway after only 3,500 years; i.e., 1,500–2,000 years earlier than in Fig. 3. After 3,000 years, the population has reached Gibraltar and moved throughout Spain. The 3500-year isochron shows that the population travels first along the coast of Spain, and then diffuses inland. This pattern is also supported by the radiocarbon dates, e.g., in Fig. 2 of [23]. Again the population eventually stabilises at the carrying capacity.

## *5. Discussion*

The comparison of these results with the archaeological and radiometric data on the Neolithisation of Europe (Fig. 1) shows several important points of coincidence.

1. Both models satisfactorily simulate the expansion pattern of early agricultural settlements in the Levant and Asia Minor, as implied by radiocarbon dates [8].
2. Model 2 (Fig. 4) is more successful in simulating the spread of early agricultural communities in the south-east of Europe (e.g., as reflected in the Karanovo-Starcevo-Körös-Criş cultural complex).
3. Model 2 reproduces the accelerated spread of LBK sites over the loess plains of Central Europe along the Danube–Rhine axis. (This, however, is not surprising as advection along the rivers has been enforced in this model.)
4. Model 2 provides an acceptable fit to the spread of early agricultural communities further east, into the forest-steppe areas of the East European Plain (e.g., as indicated by the early sites of Cucuteni-Tripolye). These facts may be viewed as indirect evidence of the importance of human migration from the Near East in the early spread of agriculture in South-Eastern and Central Europe. They equally suggest the major importance of the river pathways in the early agricultural colonisation of Central Europe.
5. Model 2 is successful in reproducing the agricultural colonisation of Britain and southern Scandinavia, via the crossing of narrow water barriers.
6. The model satisfactorily reproduces the agricultural colonisation of the northern Mediterranean coastal area via sea routes.

While the model simulates many large-scale features of the spread of incipient farming into Europe, as implied by the radiocarbon dates, it is by no means exhaustive or complete. There are some major considerations that have not yet been addressed.

In selecting which rivers to include, a more or less arbitrary approach was adopted (including only the Rhine and Danube, and neglecting other rivers which might also influence the propagation). Thus an attempt should be made to adopt a more quantitative approach in our selection of 'major' rivers. (As slight mitigation, it might be argued that the Danube and Rhine valleys were especially attractive to the farmers, due to the presence there of fertile loess soil.)

Our model shows an unrealistically high speed of advance through northern Africa. In reality, the Sahara forms a natural barrier to farming. The same applies to agricultural colonisation of the forested East European Plain and Finland; existing archaeological evidence suggests much later emergence of farming in these areas. Therefore, the model has to be further enriched by proper allowance for the landscape suitability for agriculture.

Our model is flexible enough to admit appropriate refinements. For example, Steele et al. [36] suggest that the broad-scale vegetation of the time is an important factor, and the carrying capacity should be allowed to depend on this. This would not change the speed of advance of the wave front $U$, as this is independent of carrying capacity in the standard model. Thus, the question arises as to whether the human mobility (and hence, the diffusivity) should also depend on the vegetation, soil and biomass of the varying regions; and if so, in what way. Alternatively, Eq. (2) could be modified to allow the carrying capacity to play a role in the propagation speed $U$, as suggested by Cohen [13].

Our model currently considers only land farmers; i.e., it includes only one type of population. Aoki et al. [6] considered a three-population model, including separate populations for the initial hunter-gatherers, for the invading farmers, and for a population that have converted from hunter-gathering to farming. Each population then has its own carrying capacity, mobility, growth rate and advective velocity, which might in turn be differently influenced by the environmental factors. One alternative refinement might include the spread of pottery making as another trait of the Neolithic, and try to model its unique signature in North-Eastern Eurasia. This may necessitate the introduction of yet another initial centre of diffusion, in this case located in the eastern steppe area [17].

Murray [28] observes that the classical diffusion used here is only applicable under certain conditions; notably it requires relatively small population densities. Long-range migration events (or nonlocal effects) can be important, especially where travel is facilitated as near rivers and sea coast. These effects will affect the diffusion term, and they are the subject of ongoing study.

Finally, our simulation of the Neolithisation of Europe — successful as it appears — might most realistically be viewed as a 'calibration' of our model. The significance of our calculations could then be tested by applying similar models to other regions.

## *Acknowledgements*

We thank Graeme Ackland for helpful discussions.## *Appendix A1*

The diffusion equation in population dynamics models can be directly deduced from population balances under a random walk of individuals. Here we derive this equation allowing for an anisotropic random walk in two dimensions. Similar one-dimensional calculations can be found, e.g., in Ref. [15]. We start from a model of the random walk in an infinite two-dimensional habitat. During each time interval $\tau$ individuals move either a distance of $\lambda_1$ to the right, $\lambda_2$ to the left, $\omega_1$ up, or $\omega_2$ down, with probabilities, $\alpha$, $\beta$, $\rho$ and $\xi$, respectively. Here $\alpha+\beta+\rho+\xi=1$. The general expression for the population density at any point $(x,y,t)$ then follows as

$$N(x,y,t) = \alpha N(x-\lambda_1, y, t-\tau) + \beta N(x+\lambda_2, y, t-\tau) + \rho N(x, y-\omega_1, t-\tau) + \xi N(x, y+\omega_2, t-\tau) \quad (A1)$$

$$N(x - \lambda_1, y, t - \tau) = N(x, y, t) - \tau \frac{\partial N(x,y,t)}{\partial t} - \lambda_1 \frac{\partial N(x,y,t)}{\partial x} + \frac{\tau^2}{2} \frac{\partial^2 N(x,y,t)}{\partial t^2}$$
$$+ \tau \lambda_1 \frac{\partial^2 N(x,y,t)}{\partial x \partial t} + \frac{\lambda_1^2}{2} \frac{\partial^2 N(x,y,t)}{\partial x^2} \quad \text{(A2a)}$$

$$N(x + \lambda_2, y, t - \tau) = N(x, y, t) - \tau \frac{\partial N(x,y,t)}{\partial t} + \lambda_2 \frac{\partial N(x,y,t)}{\partial x} + \frac{\tau^2}{2} \frac{\partial^2 N(x,y,t)}{\partial t^2}$$
$$- \tau \lambda_2 \frac{\partial^2 N(x,y,t)}{\partial x \partial t} + \frac{\lambda_2^2}{2} \frac{\partial^2 N(x,y,t)}{\partial x^2} \quad \text{(A2b)}$$

$$N(x, y - \omega_1, t - \tau) = N(x, y, t) - \tau \frac{\partial N(x,y,t)}{\partial t} - \omega_1 \frac{\partial N(x,y,t)}{\partial y} + \frac{\tau^2}{2} \frac{\partial^2 N(x,y,t)}{\partial t^2}$$
$$+ \tau \omega_1 \frac{\partial^2 N(x,y,t)}{\partial y \partial t} + \frac{\omega_1^2}{2} \frac{\partial^2 N(x,y,t)}{\partial y^2} \quad \text{(A2c)}$$

$$N(x, y + \omega_2, t - \tau) = N(x, y, t) - \tau \frac{\partial N(x,y,t)}{\partial t} + \omega_2 \frac{\partial N(x,y,t)}{\partial y} + \frac{\tau^2}{2} \frac{\partial^2 N(x,y,t)}{\partial t^2}$$
$$- \tau \omega_2 \frac{\partial^2 N(x,y,t)}{\partial y \partial t} + \frac{\omega_2^2}{2} \frac{\partial^2 N(x,y,t)}{\partial y^2} \quad \text{(A2d)}$$

Substituting Eqs. (A2a)–(A2d) into Eq. (A1) and rearranging gives

$$N(x,y,t) = (\alpha + \beta + \rho + \xi) N(x,y,t) - (\alpha + \beta + \rho + \xi)\tau \frac{\partial N(x,y,t)}{\partial t} - (\alpha \lambda_1 - \beta \lambda_2) \frac{\partial N(x,y,t)}{\partial x}$$
$$- (\rho \omega_1 - \xi \omega_2) \frac{\partial N(x,y,t)}{\partial y} + (\alpha + \beta + \rho + \xi) \frac{\tau^2}{2} \frac{\partial^2 N(x,y,t)}{\partial t^2} + (\alpha \lambda_1 - \beta \lambda_2)\tau \frac{\partial^2 N(x,y,t)}{\partial x \partial t}$$
$$+ (\rho \omega_1 - \xi \omega_2)\tau \frac{\partial^2 N(x,y,t)}{\partial y \partial t} + \frac{(\alpha \lambda_1^2 + \beta \lambda_2^2)}{2} \frac{\partial^2 N(x,y,t)}{\partial x^2} + \frac{(\rho \omega_1^2 + \xi \omega_2^2)}{2} \frac{\partial^2 N(x,y,t)}{\partial y^2}$$

Now allowing $\alpha + \beta + \rho + \xi = 1$, and assuming that all the step directions are equally probable, but their lengths depend on direction, we have $\alpha = \beta = \rho = \xi = \frac{1}{4}$, $\lambda_2 = \omega_2 = \lambda$ and $\lambda_1 = \lambda(1+\kappa)$ and $\omega_1 = \lambda(1+\sigma)$, which yields

$$N(x,y,t) = N(x,y,t) - \tau \frac{\partial N(x,y,t)}{\partial t} - \frac{\lambda \kappa}{4} \frac{\partial N(x,y,t)}{\partial x} - \frac{\lambda \sigma}{4} \frac{\partial N(x,y,t)}{\partial y}$$
$$+ \frac{\tau^2}{2} \frac{\partial^2 N(x,y,t)}{\partial t^2} + \frac{\lambda \kappa}{4} \tau \frac{\partial^2 N(x,y,t)}{\partial x \partial t} + \frac{\lambda \sigma}{4} \tau \frac{\partial^2 N(x,y,t)}{\partial y \partial t}$$
$$+ \frac{(2\lambda^2 + 2\lambda^2 \kappa + \lambda^2 \kappa^2)}{8} \frac{\partial^2 N(x,y,t)}{\partial x^2} + \frac{(2\lambda^2 + 2\lambda^2 \sigma + \lambda^2 \sigma^2)}{8} \frac{\partial^2 N(x,y,t)}{\partial y^2}$$

Now we take $\tau \to 0$ and require that, $\lambda^2/\tau$, $\lambda\kappa/\tau$ and $\lambda\sigma/\tau$ all tend to constants. This implies that

$$\lambda^2 = O(\tau), \quad \lambda\kappa = O(\tau), \quad \lambda\sigma = O(\tau), \quad \lambda^2\kappa = o(\tau) \quad \text{and} \quad \lambda^2\sigma = o(\tau).$$

Then we have λ, κ and σ ~ τ$^{1/2}$ . Hence λ, τ, κ and σ all tend to zero, certain combinations of parameters remain constant, and we introduce the following notation:

$$\lambda^2/4\tau = \nu, \quad \lambda\kappa/4\tau = V_x \text{ and } \lambda\sigma/4\tau = V_y. \tag{A3}$$

In terms of these variables, we obtain the following governing equation for the population density:

$$\frac{\partial N(x,y,t)}{\partial t} + V_x \frac{\partial N(x,y,t)}{\partial x} + V_y \frac{\partial N(x,y,t)}{\partial y} = \nu \left( \frac{\partial^2 N(x,y,t)}{\partial x^2} + \frac{\partial^2 N(x,y,t)}{\partial y^2} \right),$$

which is known as the diffusion-advection equation with the diffusivity and advection velocity given in Eq. (A3). More conveniently, we can write the advection speed as

$$|\mathbf{V}| = V = \frac{\lambda \mu}{4\tau},$$

where $\mu = \sqrt{\kappa^2 + \sigma^2}$. Here ν and $V$ are independent of the coordinate choice, and the result therefore holds on a spherical surface.

We note that, to this approximation, the diffusivity is not affected by the anisotropy of the random walk because both κ and σ must tend to zero as τ → 0.

# Figure Captions

**Figure 1.** A map of the Neolithic dispersal in Europe according to radiocarbon dates, given in years BP.

**Figure 2.** Shades of greyscales show the distribution of diffusivity, $\nu$, near the Danube estuary. White corresponds to $\nu=0.0$ and the darkest grey, to $\nu=12.5$ with a linear transition in between. The variation of $\nu$ is due to the altitude variations. The orientations of the advection velocity due to both rivers and coastlines are shown with black dashes; the direction of the velocity is determined, depending on the distribution of the population density, to be towards smaller values of $N$.

**Figure 3** An isochron map of the simulated Neolithic dispersal in Model 1. Environmental effects considered: altitude; latitude and sea travel. Simulation begins at time $t_0=0$ years and time of arrival is shown in years after $t_0$.

**Figure 4.** An isochron map of the simulated Neolithic dispersal in Model 2. Environmental effects considered: altitude; latitude; sea travel; major rivers and coastlines. Simulation begins at time $t_0=0$ years and time of arrival is shown in years after $t_0$. Note the tongue in the isochrons near the Rhine–Danube valleys, which later results in accelerated spread to Western and Eastern Europe.

# Figures

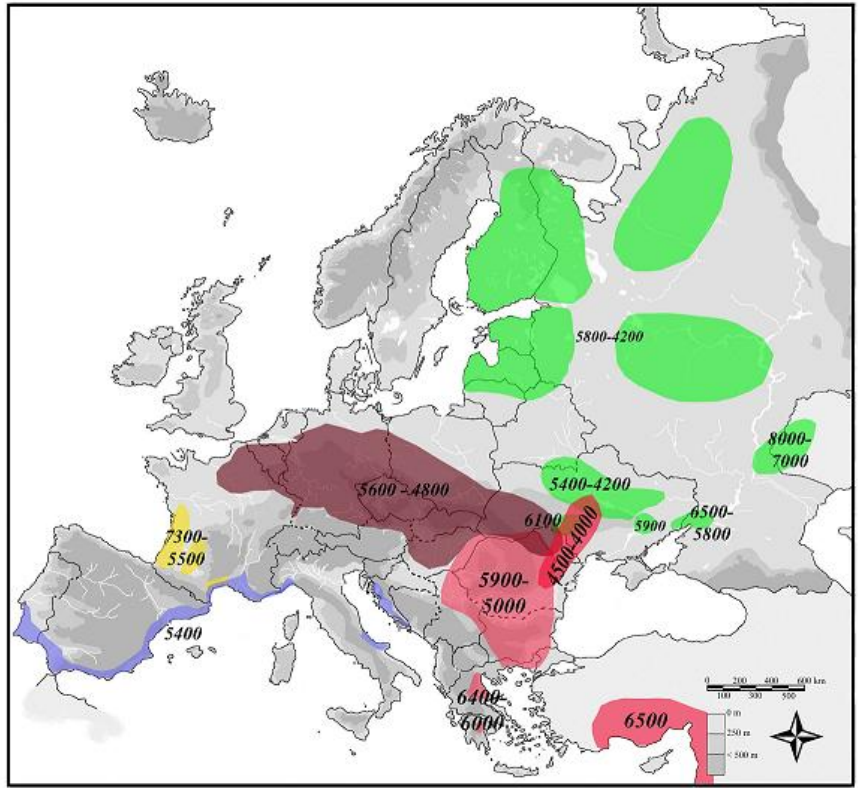

**Figure 1.** A map of the Neolithic dispersal in Europe according to radiocarbon dates, given in years BP.

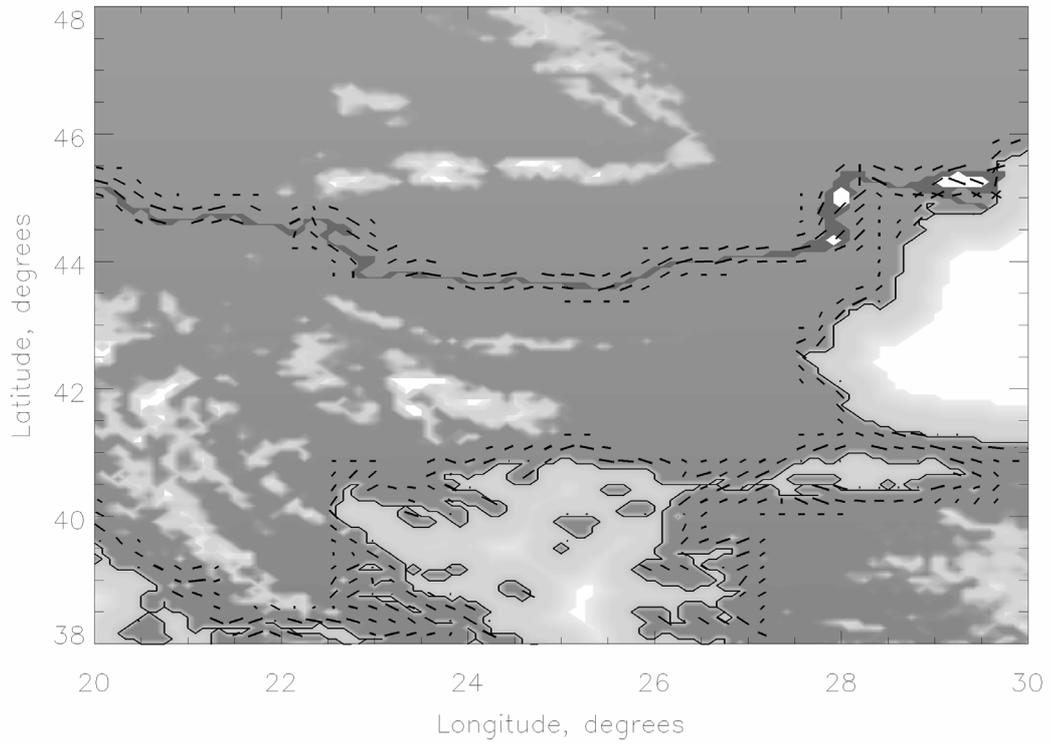

**Figure 2.** Shades of greyscales show the distribution of diffusivity, ν, near the Danube estuary. White corresponds to ν=0.0 and the darkest grey, to ν=12.5 with a linear transition in between. The variation of ν is due to the altitude variations. The orientations of the advection velocity due to both rivers and coastlines are shown with black dashes; the direction of the velocity is determined, depending on the distribution of the population density, to be towards smaller values of *N*.

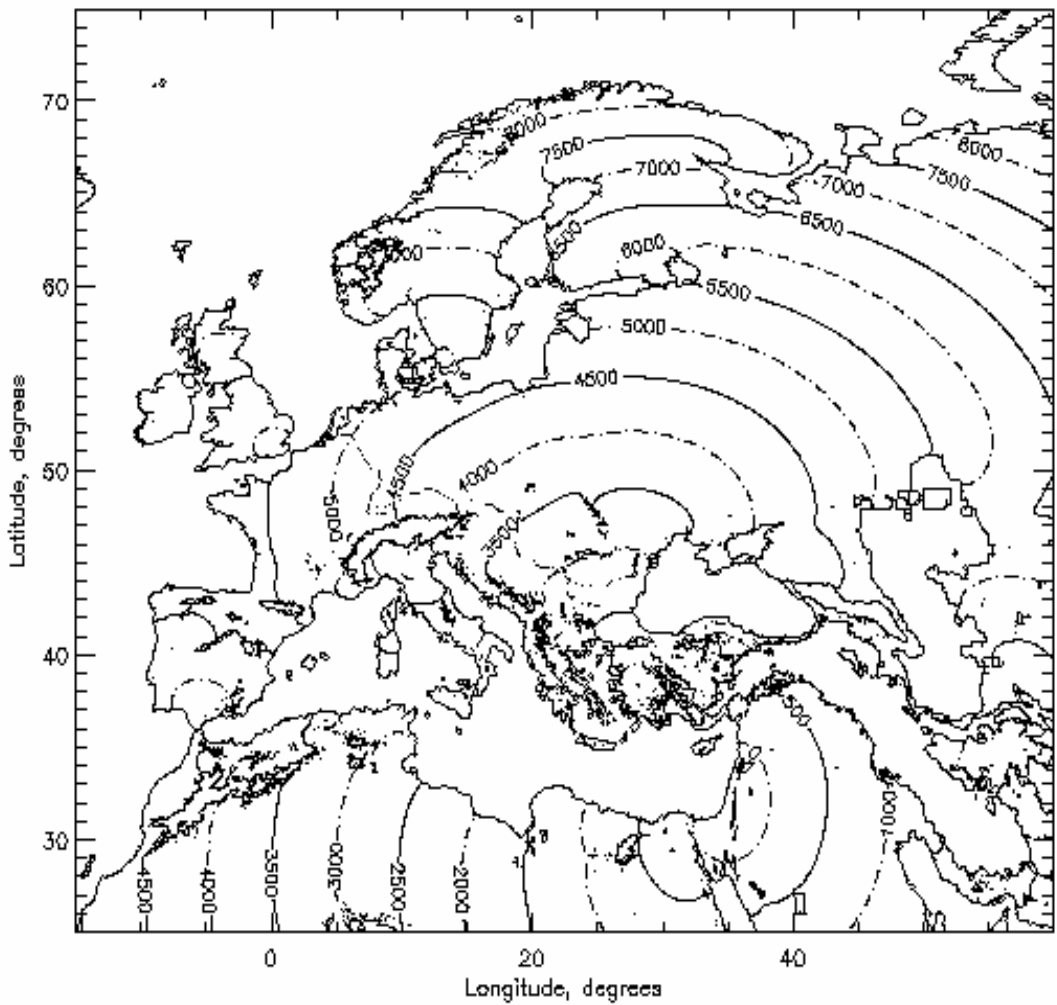

**Figure 3**   An isochron map of the simulated Neolithic dispersal in Model 1. Environmental effects considered: altitude; latitude and sea travel. Simulation begins at time $t_0$=0 years and time of arrival is shown in years after $t_0$.

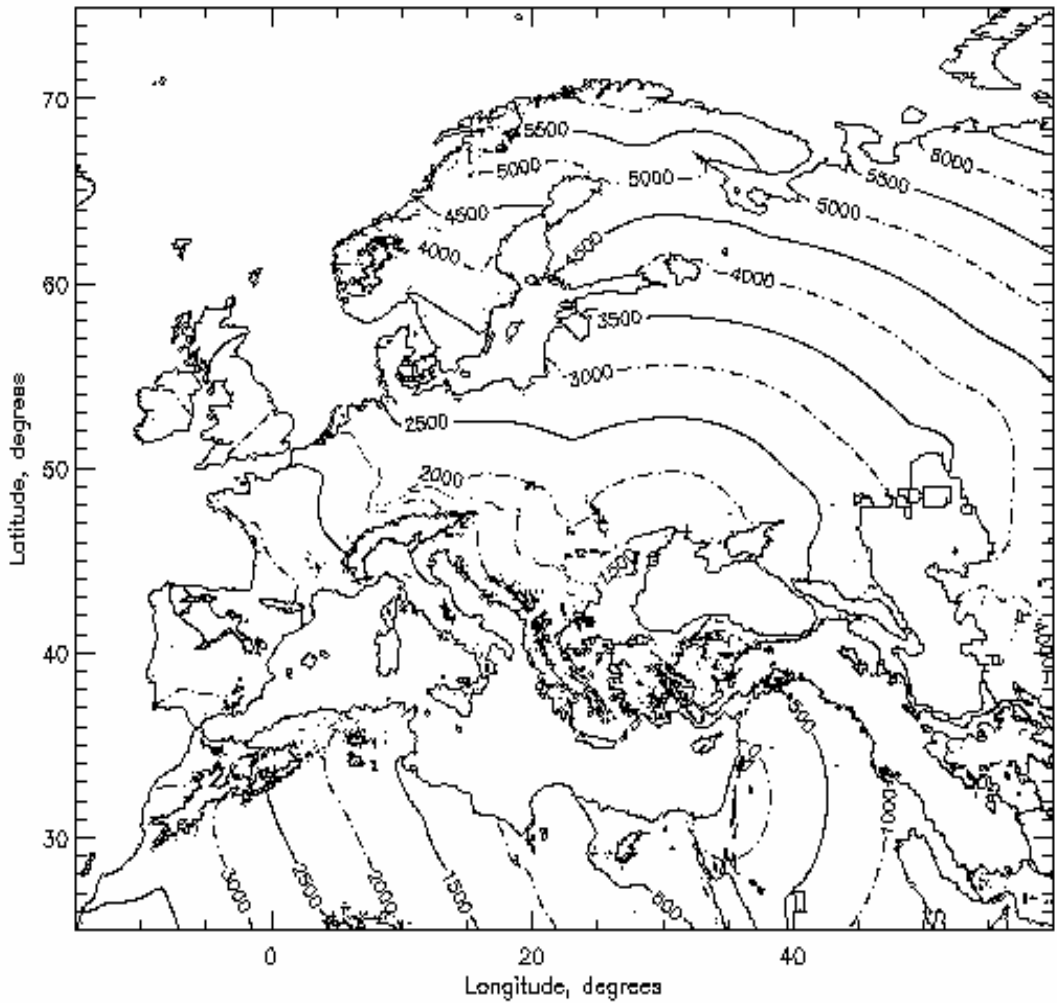

**Figure 4.** An isochron map of the simulated Neolithic dispersal in Model 2. Environmental effects considered: altitude; latitude; sea travel; major rivers and coastlines. Simulation begins at time $t_0$=0 years and time of arrival is shown in years after $t_0$. Note the tongue in the isochrons near the Rhine–Danube valleys, which later results in accelerated spread to Western and Eastern Europe.